\documentclass[useAMS,usenatbib]{mn2e}
\usepackage{graphicx}

\title[Reliability of kinematical evidence for dark matter]
{How reliable is the kinematical evidence for dark matter: the effects of
non-sphericity, substructure and streaming motions}
\author[T. Sanchis et al.]{Teresa Sanchis$^{1}$, Ewa L. {\L}okas$^{2}$ and Gary A. Mamon$^{3,4}$\\
$^1$Departament d'Astronomia i Meteorologia. Universitat de Barcelona, Avenida Diagonal 647, E-08028 Barcelona, Spain \\
$^2$Nicolaus Copernicus Astronomical Center, Bartycka 18, 00-716 Warsaw, Poland\\
$^3$Institut d'Astrophysique de Paris (CNRS UMR 7095), 98 bis Bd Arago, F-75014 Paris, France \\
$^4$GEPI (CNRS UMR 8111), Observatoire de Paris, F-92195 Meudon, France
}

\begin{document}

\maketitle

\begin{abstract}

Using cosmological $N$-body simulations of dark matter haloes, we study the
effects of non-sphericity, substructure and streaming motions in reproducing
the structure and internal kinematics of clusters of galaxies from kinematical
measurements. Fitting an NFW model to the 3D density profile, we determine the
virial mass, concentration parameter and velocity anisotropy of the haloes, and
then calculate the profiles of projected velocity moments, as they would be
measured by a distant observer. Using these mock data, we apply a Jeans
analysis for spherical objects to reproduce the line-of-sight velocity
dispersion and kurtosis and fit the three parameters. We find that the
line-of-sight velocity dispersion and kurtosis profiles of a given halo can
vary considerably with the angle of view of the observer.  We show that the
virial mass, concentration parameter and velocity anisotropy of the haloes can
be reproduced satisfactorily independently of the halo shape, although the
virial mass tends to be underestimated, the concentration parameter
overestimated, while the recovered anisotropy is typically more tangential than
the true one. The mass, concentration and velocity anisotropy of haloes are
recovered with better precision when their mean velocity profiles are near
zero.

\end{abstract}


\section{Introduction}

There is a long tradition of determining the internal kinematical properties of
bound systems based on the Jeans equations, which are velocity moments of the
collisionless Boltzmann equation, and which link the 3D velocity moments (e.g.
velocity dispersion and kurtosis) to the potential gradient. The Jeans
equations are thus used to model the mass and velocity distribution inside
elliptical galaxies \citep[e.g.][]{bm}, clusters of galaxies
\citep[e.g.][]{kg82,mer2}, as well as globular clusters  \citep[e.g.][]{mk}.
The standard Jeans approach for the determination of the mass and velocity
distribution assumes equilibrium and sphericity of the system. However, even in
the inner parts of an object, there is non-virialized matter, for example
matter falling into a cluster for the first time or smaller clumps in the
process of relaxation. Such non-virialized matter could produce misleading
results when a cluster is studied through a Jeans analysis. Moreover, clusters
of galaxies are not observed to be spherically symmetric \citep{bing,wu}, nor
are simulated structures of dark matter particles with the masses of clusters
of galaxies \citep{cl96,js02}.

One way to avoid non-virialized matter within a cluster is to restrict the
Jeans analysis to the population of elliptical galaxies, which is thought to be
dynamically relaxed \citep{TS84,bktm,lm03}. The question of how the existing
substructure and non-sphericity may affect the results can only be fully
addressed by cosmological $N$-body simulations including realistic galaxy
formation, where all 3-dimensional information would be available.

The effect of incomplete virialization of structures of dark matter particles
seen in cosmological $N$-body simulations on the estimates of the mass of a
single cluster through the Jeans equation has been addressed by \citet*{tbw}.
They showed that even for significantly perturbed haloes, the mass $M(r)$ at
distances larger than 2\% of the virial radius inferred by the proper Jeans
analysis is within 30\% (r.m.s.) of the true mass and departs from it by less
than 20\% (r.m.s.) for average or relaxed haloes.

In this work, we use cosmological $N$-body simulations and analytical modelling
to study the effect of departures from equilibrium and non-sphericity of dark
matter haloes on the inferred properties of the halo and velocity distribution
of its particles. We measure the mass and velocity distribution in the haloes
and calculate the projected velocity moments as an observer would do. We then
perform a kinematic analysis based on the Jeans equations to check to what
extent we can reproduce the properties of the haloes from the second-order
(line-of-sight velocity dispersion) and fourth-order (line-of-sight kurtosis)
velocity moments.

The paper is organized as follows. In Section~2, we describe the $N$-body
simulations used and calculate the properties of the haloes chosen for
analysis. In Section~3, we estimate projected velocity moments of the haloes.
Section~4 is devoted to analytical modelling of those moments based on Jeans
formalism and testing its reliability in reproducing the properties of the
haloes. The discussion follows in Section~5.

\section{The simulated dark matter haloes}

We have used the $N$-body simulations carried out by \citet{Hatton+03} with their {\tt
GalICS} hybrid $N$-body/semi-analytic model of hierarchical galaxy formation. The
description of this model can be found in  \citet{Hatton+03}. The $N$-body simulation
contains $256^{3}$ particles of mass $8.272 \times 10^9$ M$_{\odot}$ in a box of size
150 Mpc and its softening length amounts to spatial resolution of 29 kpc. The simulation
was run for a flat universe with cosmological parameters $\Omega_{0} = 0.333,
\Omega_{\rm \Lambda} = 0.667$, $H_{0}=66.7\,\rm km\,s^{-1}\,Mpc^{-1}$, and $\sigma_8 =
0.88$.  Once the simulation is run, haloes of dark matter are detected with a
`friends-of-friends' (FOF) algorithm \citep{DEFW85} using a variable linking length such
that the minimum mass of the FOF groups is $1.65 \times 10^{11}$ M$_{\odot}$ (20
particles) at any time step. With this method, over $2 \times 10^{4}$ haloes are
detected at the final timestep, corresponding to the present-day ($z=0$) Universe.

We restrict our analysis to dark matter particles. Although the {\tt GalICS} simulations
we use include galaxy formation and evolution, this part of the simulations is based on
a semi-analytical approach, which for our purposes is not yet satisfactory. For example,
the isotropic velocity distribution of {\tt GalICS} galaxies is imposed and is not a
result of virialization. We therefore conclude that the galaxies in these simulations
are not reliable tracers of the overall dynamical properties of the haloes and cannot be
used to infer the density and velocity distributions from the `observed' velocity
moments.

For our analysis, we have chosen the ten most massive haloes formed in the simulation
box (labelled hereafter in order of decreasing virial mass as halo 1, halo 2 and so
forth). We estimated their virial radii, $r_{\rm 100}$, as the distances from the centre
where the mean density is $100$ times the present critical density \citep[in agreement
with the so-called spherical collapse model, see][]{ks96,lh01}. The centres of the
haloes are determined as the local density maxima which turn out to be slightly
different than the centres of mass found using the FOF algorithm. Within the virial
radius, the haloes have $2 \times  10^5$ particles for the most massive halo and $4
\times 10^4$ particles for the least massive one of the haloes we have chosen. The
masses of the haloes, $M_{100} = M(r_{100})$, are listed in Table~\ref{properties}.

%%%%%%%%%%%%%%%%%%%%%%%%%TABLE%%%%%%%%%%%%%%%%%%%%%%%%%%%%%%%%%%%%%%%%%%%%%%%%%%%%%%%%%%%%
\begin{table}
\tabcolsep 2pt
\caption[]{Three-dimensional properties of the simulated haloes}
\begin{center}
\begin{tabular}{ccccccc}
\hline \hline
Halo & Axis ratio & $M_{\rm 100}$ & $r_{\rm 100}$ & $c$ &
$\langle\beta\rangle$ & $\overline{v}_{\rm r}$  \\
	&	& ($10^{14}$ M$_{\odot}$) & (Mpc) & &  &($v_{100}$)\\
\hline
1  & $2.3\,:\,1.3\,:\,1$  &  16.7$\,\ $  & 3.2  & 5.6 & $~~0.40 \pm 0.06$ &~~0.005 \\
2  & $1.9\,:\,1.1\,:\,1$  &  8.5  & 2.5  & 5.0 & $~~0.13 \pm 0.03$ &~~0.103 \\
3  & $2.8\,:\,2.0\,:\,1$  &  8.1  & 2.5  & 4.5 & $~~0.24 \pm 0.05$ &~~0.134\\
4  & $3.3\,:\,1.3\,:\,1$  &  7.0  & 2.4  & 7.1 & $~~0.14 \pm 0.06$ &$-$0.260 \\
5  & $2.0\,:\,2.0\,:\,1$  &  4.5  & 2.1  & 7.9 & $~~0.15 \pm 0.07$ &~~0.065 \\
6  & $2.6\,:\,1.4\,:\,1$  &  4.3  & 2.0  & 5.0 & $~~0.36 \pm 0.05$ &~~0.028 \\
7  & $1.8\,:\,1.4\,:\,1$  &  4.1  & 2.0  & 8.9 &$-0.03 \pm 0.05$ &$-$0.079 \\
8  & $2.4\,:\,1.6\,:\,1$  &  3.9  & 2.0  &10.0~\, & $~~0.44 \pm 0.03$ &$-$0.018 \\
9  & $2.2\,:\,1.5\,:\,1$  &  3.2  & 1.8  & 7.1 & $~~0.08 \pm 0.12$ &$-$0.012 \\
10 & $1.8\,:\,1.3\,:\,1$  &  2.9  & 1.8  &10.0~\, &$-0.23 \pm 0.08$ &$-$0.168 \\

\hline
\end{tabular}
\end{center}
\label{properties}
\end{table}
%%%%%%%%%%%%%%%%%%%%%%%%%%%%%%%%%%%%%%%%%%%%%%%%%%%%%%%%%%%%%%%%%%%%%%%%%%%%%%%%%%%%%%%%

\begin{figure}
\centering
\resizebox{\hsize}{!}{\includegraphics{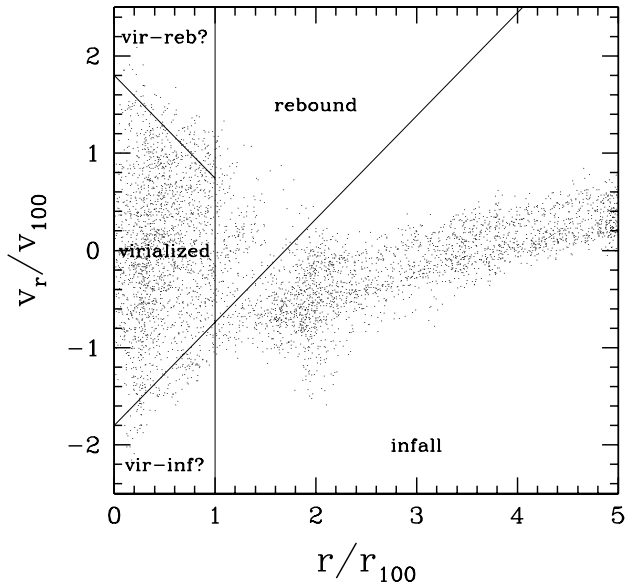}}

\caption[]{ 3D radial phase space diagram (radial velocity, Hubble flow
included, versus radial distance, both with respect to the cluster centre and
normalized to virial quantities $v_{100}$ and $r_{100}$, respectively) for a
random set of 1\% of particles of the most massive of the simulated haloes
(halo 1). The oblique lines separating the dynamical regimes were drawn `by
eye' and correspond to $v_{r}/v_{100}=1.8-1.06 \,r/r_{100}$ (\emph{bottom}) and
$v_{r}/v_{100}=-1.8+1.06\,r/r_{100}$ (\emph{top}).
}

\label{phase}
\end{figure}

All haloes have similar 3D radial phase space distributions. An example of such a
distribution for the most massive halo is presented in Figure~\ref{phase}. The Figure
shows the radial velocity (with respect to the centre of the halo) of the particles in
units of the circular velocity at $r_{\rm 100}$ ($v_{\rm 100} = v_c(r_{\rm 100}) =
\sqrt{G M_{\rm 100}/r_{\rm 100}}$) as a function of radial distance measured in units of
$r_{\rm 100}$. We have divided the phase space into different regions corresponding to
different dynamical states that we expect to find in a typical dark matter halo. Inside
the virial radius, we can identify three regions. The region with low to moderate
absolute velocities is likely to be populated by virialized particles, although there
are indications of a group of particles at $r = 0.3\,r_{100}$ slowly moving into the
halo core. On the outskirts of the velocity distribution we can find particles whose
dynamical state is not clear, as they could be either high-velocity outliers of the
virialized component or else infalling towards the cluster core or already in a rebound
regime after a passage through the centre. Beyond the virial radius, we can also find
particles in a rebound trajectory and particles on the infalling branch (which includes
particles expanding away from the cluster beyond the turnaround radius at 2.5--4 virial
radii). In Figure~\ref{phase}, these five different subsamples are denoted by {\tt
virialized}, {\tt rebound}, {\tt infall}, {\tt vir-inf?} and {\tt vir-reb?},
respectively, where the question marks indicate the uncertainty in the actual dynamical
state of these subsamples.

It is important to note that when studying projected quantities we need to consider all
five subsamples, as well as the whole sample (which will be marked hereafter by `all'),
because, as seen projected on the sky, there are particles belonging to each subsample
that fall inside the `virial' cylinder whose projected radius is the virial radius
$r_{100}$, but there is no way of determining which of these particles are actually
within the `virial' sphere of radius $r_{100}$ (inscribed in the virial cylinder).

\begin{figure}
\centering
\resizebox{\hsize}{!}{\includegraphics{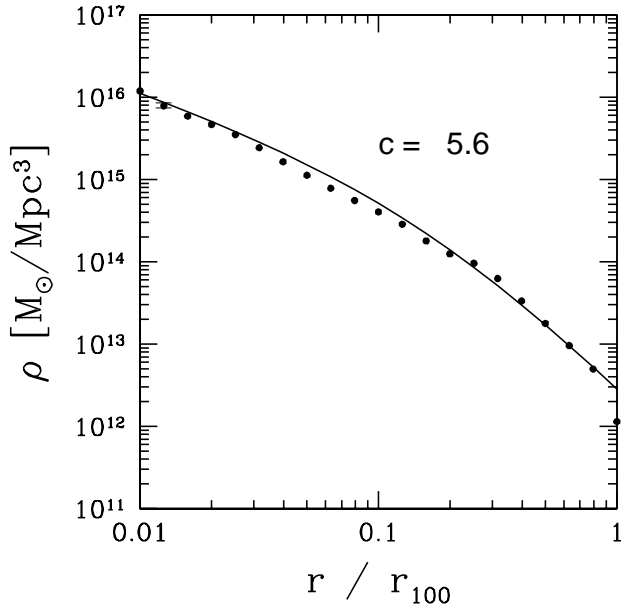}}

\caption[]{Density profile of halo 1. The measurements were done in radial bins
of equal logarithmic length and errors were estimated as Poisson fluctuations.
The error bar in the second point shows the maximum error assigned to
measurements. The \emph{curve} is the best-fitting NFW profile with
concentration $c=5.6$.
}

\label{profile}
\end{figure}

Figure~\ref{profile} shows the density distribution in the most massive halo 1
as a function of radial distance in units of $r_{\rm 100}$. We found that the
measured density profile is well approximated by the NFW formula \citep*{nfw}

\begin{equation}    \label{ro}
      \frac{\varrho(s)}{\varrho_{c,0}} = \frac{\Delta_c \,c^2 g(c)}{3 \,s\,(1+ c
     s)^2} \ ,
\end{equation}

\noindent
where $s=r/r_{\rm 100}$, $\varrho_{c,0}$ is the present critical density,
$\Delta_c=100$, $c$ is the concentration parameter and $g(c) = [\ln (1+c) -
c/(1+c)]^{-1}$. The statistical errors are much smaller than the departures due
to substructure in the halo, but the overall fit is satisfactory. We find that
the best-fitting concentration parameter for halo 1 is $c=5.6$. The virial mass
of halo 1 is $M_{\rm 100}= 1.67 \times 10^{15}$ M$_{\odot}$  so the
concentration we estimated is consistent with the dependence of $c$ on mass
inferred from $N$-body simulations by \citet{bu}, also run with a $\Lambda$CDM
cosmology. Similar results are obtained for other haloes. The fitted parameters
are summarized in Table~\ref{properties}.

\begin{figure}
\centering
\resizebox{\hsize}{!}{\includegraphics{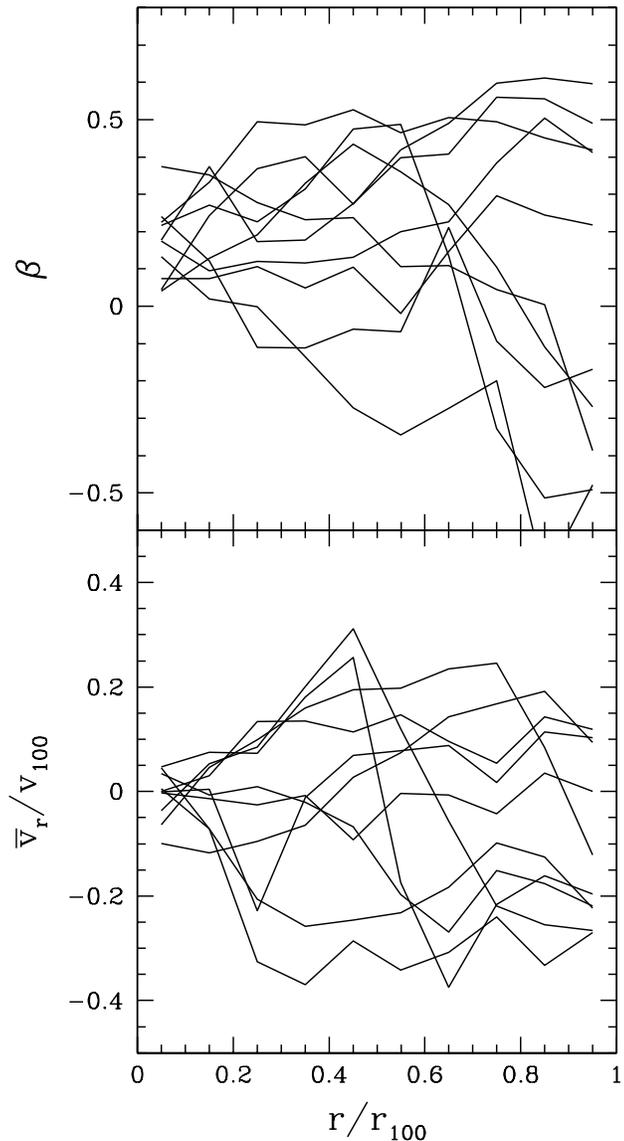}}

\caption[]{The radial profile of the anisotropy parameter $\beta$ (\emph{top
panel})  and the mean radial velocity in units of the circular velocity at
$r_{100}$ (\emph{bottom panel}) for all particles inside $r_{\rm 100}$ for the 10
haloes of Table~\ref{properties}.
}

\label{beta}
\end{figure}

In Figure~\ref{beta} we present radial profiles of the mean radial velocity in
units of the circular velocity at $r_{100}$ and the anisotropy parameter
\begin{equation}
\label{d7}
\beta=1-\frac{\sigma_\theta^2(r)}{\sigma_r^2(r)}
\end{equation}
for the ten haloes, where $\sigma_\theta$ and $\sigma_r$ are the velocity dispersions
(with respect to the mean velocities) discussed in detail in the next Sections. We show
measurements for all particles inside the sphere of radius $r_{\rm 100}$. The curves
represent the anisotropy or mean radial velocity  of dark matter particles enclosed in
shells of thickness $0.1 r_{100}$ centered at $0.05 r_{100}$, $0.15 r_{100}$ etc. As we
can see, the orbits of most of the haloes are mildly radial, with positive mean $\beta$.
We have calculated the unweighted mean anisotropy inside the virial radius. These values
are listed in Table~\ref{properties}, where the uncertainties are the dispersions of the
values about the unweighted mean. The anisotropy does not vary strongly with radius so
in the modelling which follows we will assume it to be constant and equal to the mean
value inside the virial radius. However, it is interesting to note that the variation of
$\beta$ with distance from the centre is very different for each of the analyzed haloes.
As for the mean radial velocity, we can see from Figure~\ref{beta} that it is consistent
on average with zero inside $r_{100}$ for most haloes, albeit with non-negligible radial
variations caused by internal streaming motions, including some abrupt variations
suggesting the presence of smaller haloes (akin to groups of galaxies) falling in or
bouncing out of the halo. In Table~\ref{properties} we list this quantity in units of
$v_{100}$. The negative sign indicates infall motion towards the centre of the halo.

Since we wish to study the effect of non-sphericity of haloes by choosing different
directions of observation, we have determined the principal axes of our haloes from
their moments of inertia, using particles at radial distances up to $r_{\rm 100}$. The
ratios of the eigenvalues of the inertia tensor, listed in Table~\ref{properties}, show
significant departures from sphericity.

\section{Velocity moments of dark matter particles}

We now study the kinematical properties of haloes as they would be seen by a distant
observer. The quantities discussed are all projected along the line of sight. With the
Jeans formalism, we can only model the projected velocity moments of objects in
equilibrium, and thus we cannot make any prediction about the velocity moments of the
regions of phase space called  `rebound', `infall', `vir-reb?' and `vir-inf?' in
Figure~\ref{phase}. Moreover, not all particles inside $r_{\rm 100}$ are virialized, as
they could be falling into the core for the first time or be on a rebound orbit not yet
in equilibrium. However, the division of the phase space into the 5 regions
(Fig.~\ref{phase}) was made by eye, with no exact determination of the dynamical state
of the particles, and it was only based on the radial component of the particle
velocities. This could produce misleading results in the analysis of the velocity
moments, which involve all components of the velocity. For these reasons, in what
follows, we will restrict ourselves to the study of the velocity moments of all
particles inside the virial sphere of radius $r_{\rm 100}$ and all particles within the
virial cylinder of projected radius smaller than $r_{100}$. The latter would be the ones
used in the Jeans formalism by an observer unable to distinguish which particles
actually lie in the virial sphere of radius $r_{\rm 100}$.

\begin{figure}
\centering
\resizebox{\hsize}{!}{\includegraphics{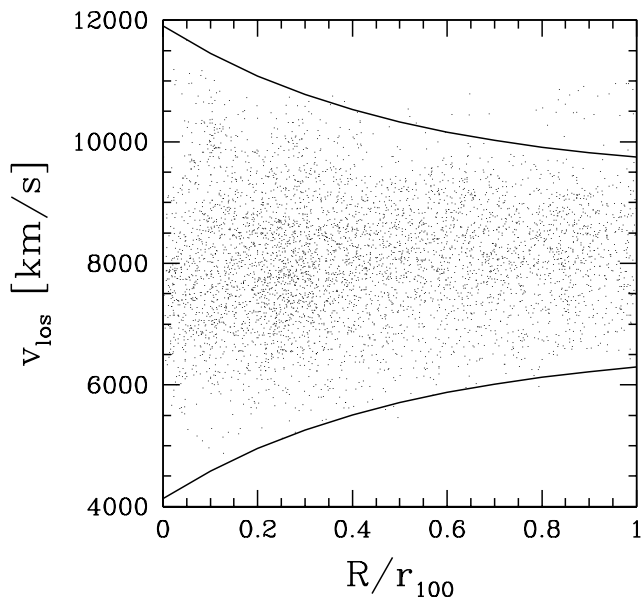}}

\caption[]{Line-of-sight velocities of a random set of 2\% of the dark matter
particles in halo 1 projected along the major axis as a function of projected
distance from the centre in units of $r_{\rm 100}$. The \emph{curves} indicate
the cuts used to distinguish between particles belonging to the halo from the
background. The vertical clumps indicate groups of particles (nearly all
within the virial sphere).
}

\label{losvel}
\end{figure}

We mimic the observations as follows. For each halo, we place an observer at either
$0^\circ$, $45^\circ$ or $90^\circ$ with respect to the major axis (see
Table~\ref{properties}) so that the three chosen directions are in the same plane
defined by the major axis and the sum of the other two axes. The choice of axes is
dictated by the non-sphericity of the set of particles within the virial sphere, but
also of the filaments of groups and other matter falling into the virial sphere out to
scales $\simeq 10\,r_{100}$. It turns out that the principal axes on scales of $\simeq
10\,r_{100}$ are very similar to those computed for $r < r_{100}$, so we restrict
ourselves to the principal axes obtained from the tensor of inertia of the particles
within the virial sphere. We then project all the particle velocities along the line of
sight and the distances on the surface of the sky. In Figure~\ref{losvel}, we show
line-of-sight velocities of dark matter particles in halo 1 projected along the major
axis as a function of projected distance from the centre in units of $r_{\rm 100}$.

Observers remove a fraction of the interlopers of a cluster by excluding the
high-velocity outliers. Here, we remove the high-velocity outliers from our mock samples
in a similar way as done by \citet{kg82} and \citet{lm03} for clusters of galaxies. The
velocity cuts are shown as solid curves in Figure~\ref{losvel}. Less than 2\% of all
particles within the virial cylinder were removed in this fashion. Although the main
body of the halo is quite well defined in velocity space, with much more particles than
there are galaxies in a cluster, the gaps between the halo and the background are not as
visible as found by \citet{lm03} for the Coma cluster. After applying this selection
procedure, the fraction of particles lying inside the cylinder of projected radius
$r_{\rm 100}$ that are actually outside the sphere of radius $r_{100}$ is between 6\%
and 35\%, with a mean of 15\%.

\begin{figure}
\centering
\resizebox{\hsize}{!}{\includegraphics{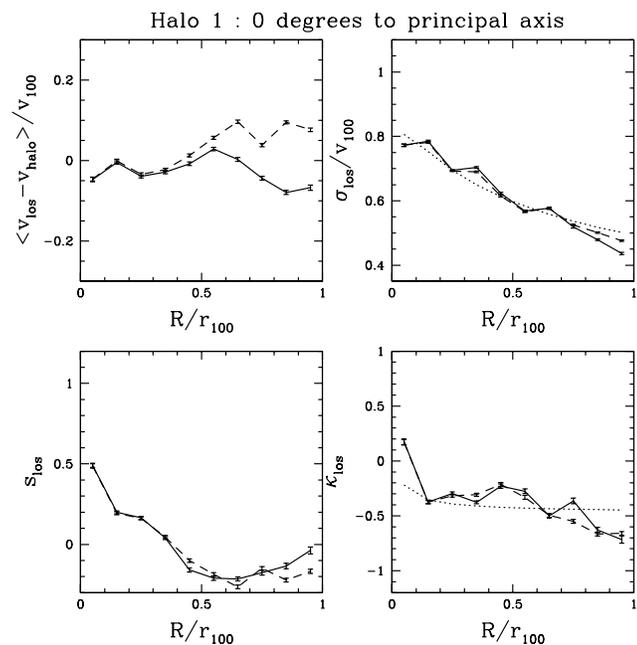}}

\caption[]{Projected velocity moments of the dark matter particles in halo 1
measured at $0^\circ$ to the principal axis. The \emph{upper left panel} shows the
mean line-of-sight velocity with respect to the velocity of the centre of the
halo in units of the circular velocity at $r_{\rm 100}$. The \emph{upper right panel}
gives the line-of-sight velocity dispersion in the same units. The two \emph{lower
panels} give the skewness (\emph{left}) and kurtosis (\emph{right}). In each panel the \emph{solid
line} shows results for particles lying inside the sphere of radius $r_{\rm
100}$, while the \emph{dashed line} is for all particles. The \emph{dotted curve} shows the fits
obtained from the Jeans equations.
}

\label{mom0}
\end{figure}

\begin{figure}
\centering
\resizebox{\hsize}{!}{\includegraphics{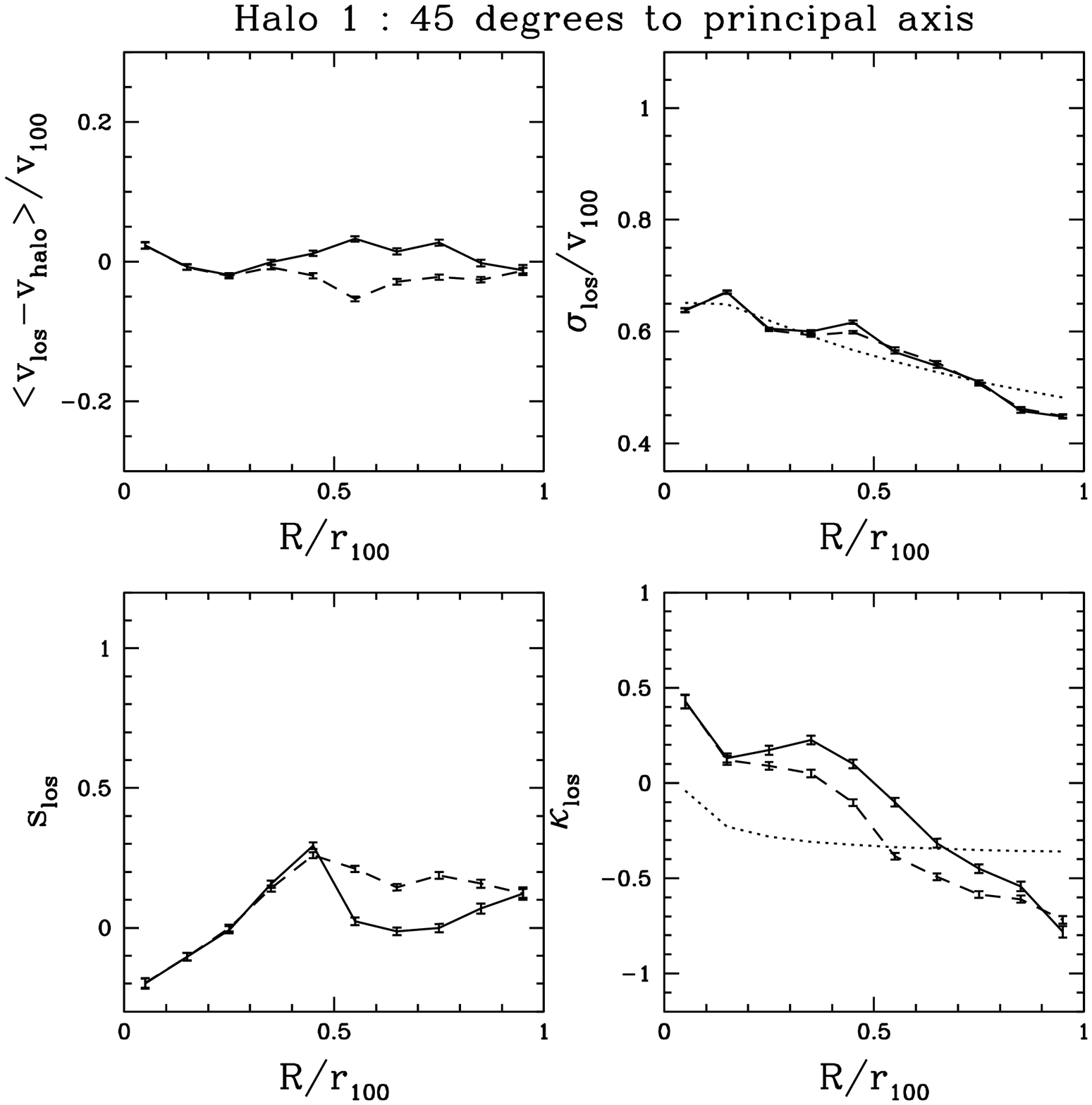}}

\caption[]{Same as Figure~\ref{mom0}, but measured at $45^\circ$ to the
principal axis.
}

\label{mom45}
\end{figure}

\begin{figure}
\centering
\resizebox{\hsize}{!}{\includegraphics{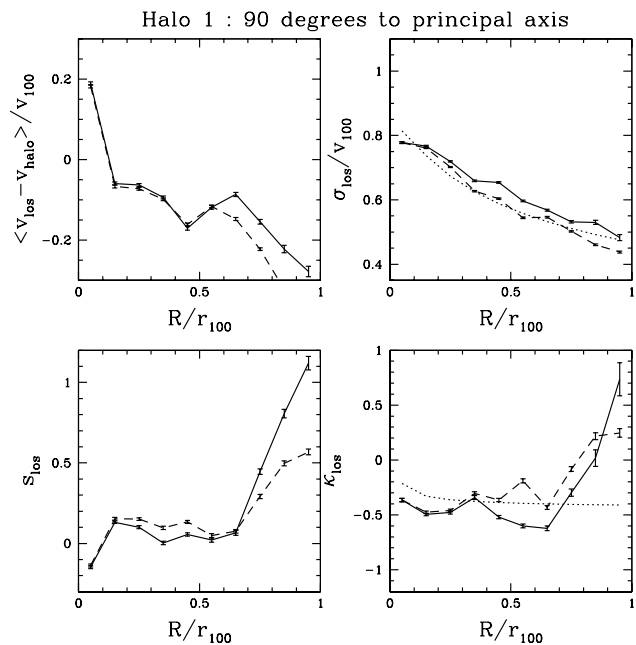}}

\caption[]{Same as Figure~\ref{mom0}, but measured at $90^\circ$ to the
principal axis.
}

\label{mom90}
\end{figure}

We divide the projected radius in ten bins and calculate, in each bin of
projected radius, the mean, dispersion, skewness, and kurtosis of the
line-of-sight velocities, $v_i$,  according to the following formulae

\begin{equation} \label{mean}
	\overline{v}_{\rm los} = \frac{1}{N} \sum_{i=1}^{N} v_i
\end{equation}
\begin{equation} \label{sig}
	\sigma_{\rm los}^{2} = \frac{1}{N-1} \sum_{i=1}^{N}(v_i
	-\overline{v}_{\rm los})^{2}
\end{equation}
\begin{equation} \label{skew}
	s_{\rm los} =  \frac{1}{N} \sum_{i=1}^{N} \left (\frac{v_i-
	\overline{v}_{\rm los}}{\sigma_{\rm los}} \right)^{3}
\end{equation}
\begin{equation} \label{kurt}
	\kappa_{\rm los} = \frac{1}{N} \sum_{i=1}^{N} \left (\frac{v_i-
	\overline{v}_{\rm los}}{\sigma_{\rm los}} \right)^{4}  -3
\end{equation}

\noindent
where $N$ represents the number of particles per bin.

To see the effects of non-sphericity, we show the results for haloes 1, 4 and 5, which
have three different shapes: halo 5 is oblate, halo 4 is roughly prolate and halo 1 is
triaxial. The results for halo 1 are shown in Figures~\ref{mom0}--\ref{mom90} for
observers situated at $0^\circ$, $45^\circ$ and $90^\circ$ with respect to the major
axis. The errors were estimated using bootstraps, but since the number of particles in
each bin is very large (of the order of $10^4$), the errors are small and do not account
for the variability of the profiles, which is mainly due to substructure. In each panel,
the dashed line shows results for all particles that in projection end up with a
projected radius smaller than the virial radius, whether or not they are actually within
the virial radius in 3D space. The solid line shows results for the particles really
lying inside $r_{\rm 100}$. In the right-hand panels presenting even moments (velocity
dispersion and kurtosis) in Figures~\ref{mom0}--\ref{mom90} we also show dotted lines
resulting from the fitting procedure based on the Jeans formalism presented in
Section~4.

The variation of the mean velocity with projected radius provides an indication
of the amount of substructure present in the halo. For observers at $0^\circ$
and $45^\circ$ with respect to the major axis of halo 1, the mean velocity with
respect to the centre of the halo is approximately zero for every radial bin,
indicating, if not the lack of substructure, at least the compensation of
effects of different substructures. This is true for the two subsets of
particles studied, especially for particles actually within the virial sphere.
For an observer at $90^\circ$ with respect to the major axis of halo 1,  we
find a departure of the mean velocity from the velocity of the centre of the
halo, which indicates the presence of substructure. As could be expected, this
radial variation of the mean velocity is more pronounced for the particles we
find in projection inside $r_{\rm 100}$ that are not necessarily within the
virial sphere of radius $r_{100}$.

\begin{figure}
\centering
\resizebox{\hsize}{!}{\includegraphics{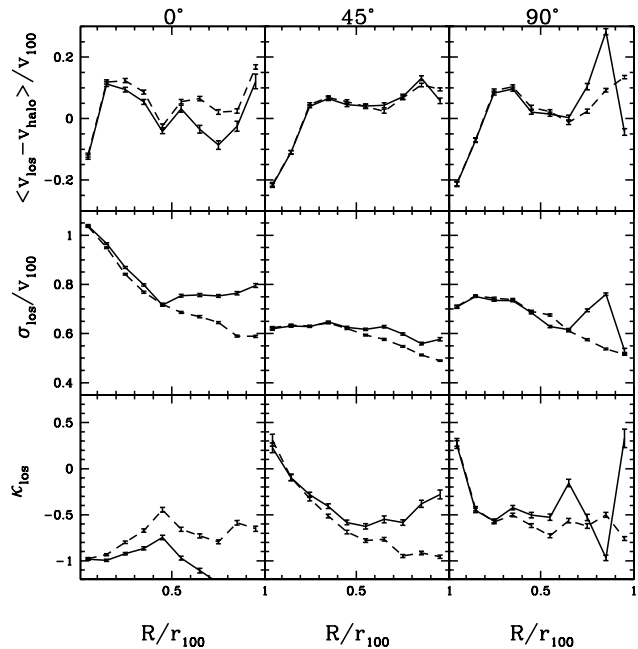}}
\caption[]{Same as Figures~\ref{mom0}-\ref{mom90} but for halo 4.
Line-of-sight skewness is not plotted. No theoretical predictions are shown.
}
\label{momrest2}
\end{figure}

\begin{figure}
\centering
\resizebox{\hsize}{!}{\includegraphics{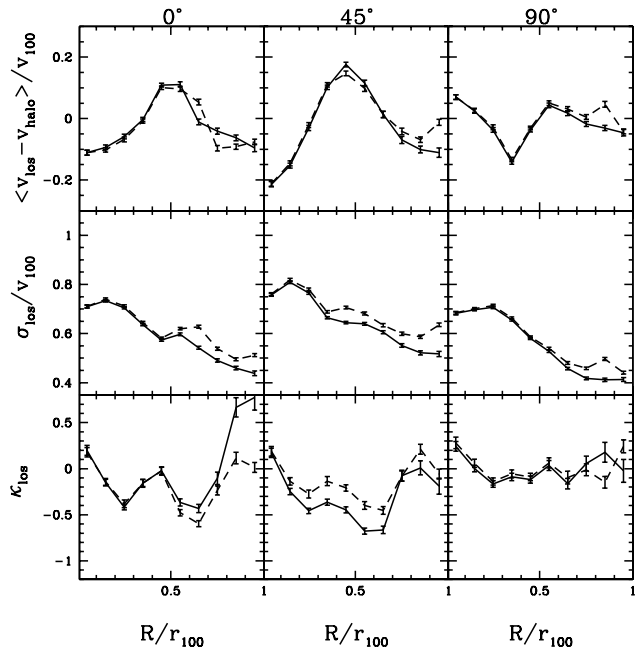}}
\caption[]{Same as Figure~\ref{momrest2}, but for halo 5.
}
\label{momrest3}
\end{figure}

Contrary to the case of elliptical galaxies, where velocity moments are
measured from spectra obtained in slits, e.g. along the major axis,  and thus
do not involve averaging in shells of similar projected radius, here and in the
analysis of galaxy motions in clusters the mean projected velocity is less
affected by the presence of global motions like rotation or infall. Therefore,
the estimation of the bulk velocities requires the 3D information. We find that
the mean velocities in the radial and tangential directions inside the virial
radius are typically of the order of few percent of $v_{\rm 100}$, except for
halo 4 and halo 10 where significant net radial infall seems to be present (see
Table~\ref{properties}).

Figures~\ref{momrest2} and~\ref{momrest3} illustrate how the velocity moments depend on
the shape of the given halo. To save space, we only show the even moments that we will
model in the next Section and the mean line-of-sight velocity with respect to the
velocity of the centre to test the relaxation of the haloes. For the sake of clarity of
the plots, we do not show the fits based on the Jeans equations, although we will
discuss them in the following section.

Even with significant noise from substructure, such as the trend for high mean velocity
at $R \simeq 0.5\,r_{100}$ for halo 5 (Fig. 9) probably arising from a group bouncing
out of the halo, we can see some common trends in Figures~\ref{mom0}--\ref{momrest3}.
First, the line-of-sight velocity dispersion and kurtosis profiles can differ
substantially for a given halo seen along three different axes (especially in halo 4).
Moreover, the velocity moments are not strongly affected by the presence of particles
outside the virial radius, i.e. the solid and dashed lines typically do not differ
significantly. This is understandable as the virial sample is a subsample of the `all'
sample and the mean difference in the number of particles is less than 20\% as
previously noted. The discrepancies between the velocity moments are more important for
larger projected distances from the centre of a halo, as the surface density of the 3D
haloes decreases faster than that of the surrounding material.

\section{Modelling of the velocity moments}

We now briefly describe the Jeans formalism for modelling the projected velocity moments
of virialized objects and apply it to recover the properties of the haloes. The detailed
description of the calculations involved can be found in \citet{lo} and \citet{lm03}
\citep[see also][]{mk,vdm}.

Our purpose here is to reproduce the projected velocity moments discussed in the
previous Section by solving the Jeans equations for the second and fourth velocity
moments and adjusting the parameters describing the mass and velocity distribution in
the haloes. We will then verify whether the fitted parameters match the real properties
of the haloes. The Jeans analysis assumes that the system is spherically symmetric and
in equilibrium, that there are no net streaming motions (no infall and no rotation) so
that the odd velocity moments vanish. As we have seen in the previous Section, none of
these is exactly the case for dark matter haloes. We want to check to what extent
violating these assumptions affects the recovered properties of the haloes.

One difference with respect to the analysis of galaxies is that the dark matter
particles are very numerous in our simulations (of order $10^5$ per halo) while the
number of galaxies in a cluster usually does not exceed a thousand. This is the reason
why the errors due to substructure will be more significant here than sampling errors
which are the dominant ones in measured velocity moments of galaxies.

The second order velocity moments are $\overline{v_r^2}$ and
$\overline{v_\theta^2}=\overline{v_\phi^2}$ and we will denote them hereafter
by $\sigma_r^2$ and $\sigma_\theta^2$ respectively. They can be calculated from
the lowest order Jeans equation \citep[e.g.][]{bm}
\begin{equation}    \label{m1}
    \frac{\rm d}{{\rm d} r}  (\nu \sigma_r^2) + \frac{2 \beta}{r} \nu
	\sigma_r^2 = - \nu \frac{{\rm d} \Phi}{{\rm d} r} \ ,
\end{equation}
where $\nu$ is the 3D density distribution of the tracer population and $\Phi$
is the gravitational potential. Since in our case dark matter particles trace
their own gravitational potential, we have $\nu(r) = \varrho(r)$. We assume
that the dark matter distribution is given by the NFW profile (\ref{ro})
characterized by its virial mass $M_{\rm 100}$ and concentration $c$. We solve
equation (\ref{m1}) assuming the anisotropy parameter of equation (\ref{d7}) to
be constant with $-\infty < \beta \le 1$. This model covers all interesting
possibilities from radial orbits ($\beta=1$) to isotropy ($\beta=0$) and
circular orbits ($\beta \rightarrow - \infty$).

The solution of the lowest order Jeans equation with the boundary condition
$\sigma_r \rightarrow 0$ at $r \rightarrow \infty$ for $\beta={\rm const}$ is
\citep[e.g.][]{lm03}
\begin{equation}	\label{m4b}
	\nu \sigma_r^2 (\beta={\rm const})= r^{-2 \beta}
	\int_r^\infty r^{2 \beta} \nu \frac{{\rm d} \Phi}{{\rm d} r} \ {\rm d}r
\ .
\end{equation}
As discussed in the previous Section, the quantity an observer would measure is
the line-of-sight velocity dispersion obtained from the 3D velocity dispersion
by integrating along the line of sight \citep{bm}
\begin{equation}    \label{m3}
    \sigma_{\rm los}^2 (R) = \frac{2}{I(R)} \int_{R}^{\infty}
    \left( 1-\beta \frac{R^2}{r^2} \right) \frac{\nu \,
    \sigma_r^2 \,r}{\sqrt{r^2 - R^2}} \,{\rm d} r \ ,
\end{equation}
where $I(R)$ is the surface distribution of the tracer and $R$ is the projected
radius. In our case $I(R)$ is given by the projection of the NFW profile
\citep[see][]{lm01}. Introducing equation (\ref{m4b}) into equation
(\ref{m3}) and inverting the order of integration, the calculation of
$\sigma_{\rm los}$ can be reduced to one-dimensional numerical integration of a
formula involving special functions for arbitrary $\beta = \rm const$.

It has been established that systems with different densities and velocity
anisotropies can produce identical $\sigma_{\rm los} (R)$ profiles  \citep[see
e.g.][]{mk,mer2}. This degeneracy can be partially lifted through the modelling
of the fourth-order moment. With $\beta =\rm const$, the solution of the Jeans
equation for the fourth-order moment
\begin{equation}    \label{d11}
        \frac{\rm d}{{\rm d} r}  (\nu \overline{v_r^4}) + \frac{2 \beta}{r} \nu
	\overline{v_r^4} + 3 \nu \sigma_r^2 \frac{{\rm d} \Phi}{{\rm d} r} =0 \ ,
\end{equation}
is \citep[see][]{lo,lm03}
\begin{equation}	\label{d12}
	\nu \overline{v_r^4} (\beta={\rm const})= 3 r^{-2 \beta}
	\int_r^\infty r^{2 \beta} \nu \sigma_r^2 (r)
	\frac{{\rm d} \Phi}{{\rm d} r} \ {\rm d} r \ .
\end{equation}
By projection, we obtain the line-of-sight fourth moment
\begin{equation}         \label{d13}
    \overline{v_{\rm los}^4} (R) = \frac{2}{I(R)} \int_{R}^{\infty}
    \frac{\nu \,  \overline{v_r^4} \,r}{\sqrt{r^2 - R^2}} \ g(r, R, \beta)
    \,{\rm d} r \ ,
\end{equation}
where $g(r, R, \beta) = 1 - 2 \beta R^2/r^2 + \beta(1+\beta) R^4/(2 r^4)$.

Introducing equations (\ref{m4b}) and (\ref{d12}) into (\ref{d13}) and
inverting the order of integration, the calculation can be reduced to a double
integral. In the following, we use the fourth projected moment scaled with
$\sigma_{\rm los}^4$ in the form of projected kurtosis
\begin{equation}	\label{d14}
	\kappa_{\rm los} (R) = \frac{\overline{v_{\rm los}^4} (R)}
	{\sigma_{\rm los}^4 (R)} - 3 ,
\end{equation}
where the value of $\kappa_{\rm los} (R)=3$ valid for a Gaussian distribution
has been subtracted.

\begin{figure}
\centering
\resizebox{\hsize}{!}{\includegraphics{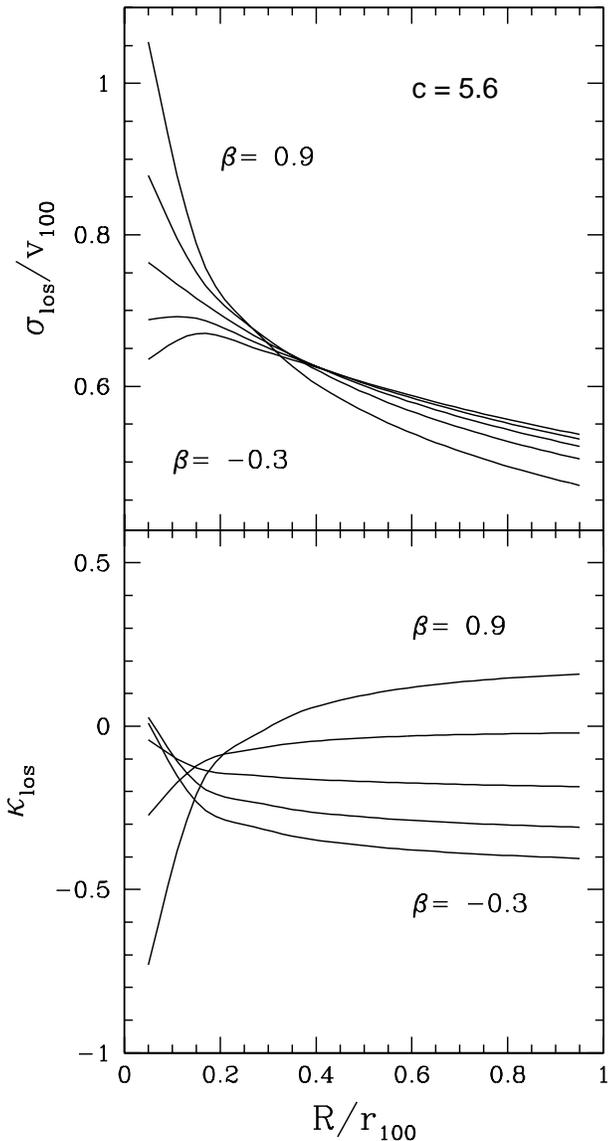}}
\caption[]{Predicted profiles of $\sigma_{\rm los} (R)$ and $\kappa_{\rm los}
(R)$ for dark matter halo with concentration $c=5.6$ and different values of
$\beta=-0.3, 0, 0.3, 0.6$ and $0.9$. The values of $\sigma_{\rm los}$ are
scaled with circular velocity at $r_{\rm 100}$ and distances are in units of
$r_{\rm 100}$. These theoretical profiles are independent of the virial mass
$M_{100}$.
}
\label{vdkurt}
\end{figure}

We can now calculate the predictions of equations (\ref{m3}) and (\ref{d13}) for a given
mass distribution and velocity anisotropy $\beta$. As already discussed, we assume that
the mass is given by the NFW distribution. For halo 1 we found the concentration of
$c=5.6$. For this value and different $\beta$ we obtain the profiles of $\sigma_{\rm
los} (R)$ and $\kappa_{\rm los} (R)$ shown in Figure~\ref{vdkurt}. The values of
$\sigma_{\rm los}$ are expressed in units of $v_{\rm 100}$ and distances are in units of
$r_{\rm 100}$. With these scalings, the predictions do not explicitly depend on the mass
of the halo. The lines show results for different values of $\beta=-0.3, 0, 0.3, 0.6$
and $0.9$, as indicated.

We see that for increasingly radial orbits (increasing $\beta$), the profile of
$\sigma_{\rm los}$ turns steeper \citep[e.g.][]{tonry}. Moreover, the kurtosis profile
becomes more convex for increasingly radial orbits as opposed to the concave shapes in
the case of isotropic and circular orbits. Since our measured kurtosis profiles in the
previous Section have a concave shape and are slightly negative we do not expect the
orbits to depart significantly from isotropic, which is consistent with the measured
anisotropy parameter (see Table~\ref{properties}).

Mimicking the procedure used by observers to infer the mass and anisotropy profiles of
galaxies and clusters, we fit the measured profiles of $\sigma_{\rm los}$ and
$\kappa_{\rm los}$ of the ten haloes by solving equations (\ref{m3}) and (\ref{d13}) and
adjusting the parameters $M_{\rm 100}$, $c$ and $\beta$, assuming that the objects are
spherical and that their dark matter distribution is given by the NFW density profile.
The fit is done by minimizing $\chi^2$ for the $20$ `data points'  of $\sigma_{\rm los}$
and $\kappa_{\rm los}$ together \citep[the data points are independent because the
number of particles in each radial bin is very large, see the discussion in the Appendix
of][]{lm03}. The data points were weighted by the assigned bootstrap errors although
they do not account for the real variability of the data and therefore the quality of
the fits in terms of $\chi^2$ is very bad. The best-fitting velocity moments found for
halo 1 are shown as dotted lines in Figures~\ref{mom0}--\ref{mom90}.

In reality, when dealing with real galaxy data for clusters, the bins would include a
few tens of objects instead of thousands, resulting in errors larger by at least an
order of magnitude. To see how the sampling errors affect the determination of the
dynamical parameters of clusters, we have also measured for each halo the velocity
moments for a set of 40 randomly chosen particles per distance bin. This number of
particles is chosen to be similar to the usual number of galaxies used in real clusters
e.g. by \citet{lm03}. Errors were assigned as described in the Appendix of \citet{lm03}.
We have then performed the same fitting procedure as described above.

\begin{figure*}
\centering
\resizebox{\hsize}{!}{\includegraphics{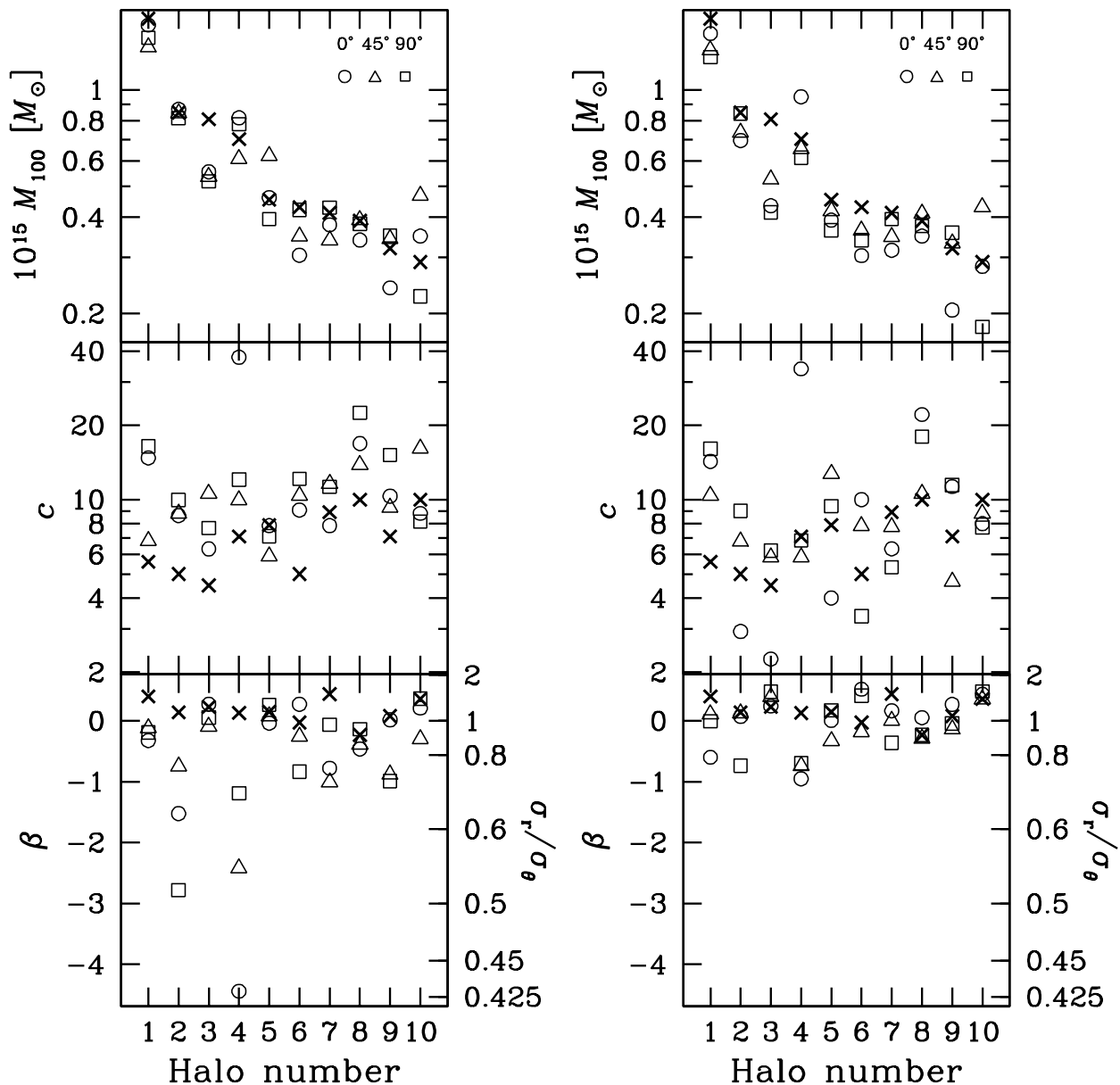}}

\caption[]{Fitted (on projected data) values of virial mass $M_{\rm 100}$,
concentration parameter $c$ and anisotropy $\beta$ of the ten haloes for the
three directions of observation with respect to the major axis of each halo:
$0^\circ$ (\emph{circles}), $45^\circ$ (\emph{triangles})
and $90^\circ$ (\emph{squares}). The
values measured on 3D data are shown with \emph{crosses}. \emph{Left:} Fitted
parameters obtained with all particles. \emph{Right:}  Fitted
parameters obtained with 40 particles per bin.
}

\label{fits}
\end{figure*}

The best-fitting parameters for the haloes estimated with all dark matter particles are
shown in the left column of Figure~\ref{fits} as different symbols depending on the
direction of observation with respect to the major axis: $0^\circ$ (circles), $45^\circ$
(triangles) and $90^\circ$ (squares). The crosses mark the `real' values of the
parameters listed in Table~\ref{properties}. In the right column of Figure~\ref{fits},
we show the best-fitting parameters estimated with 400 particles (40 particles per bin).

Due to a rather time-consuming integration involved in the calculation of kurtosis, we
restricted our analysis to the ten most massive haloes. The general conclusion is that
when taking into account all three fitting parameters, for all haloes their best-fitting
virial mass $M_{\rm 100}$ is reasonably well recovered: the discrepancy between the
best-fitting value and the real $M_{\rm 100}$ (measured using 3D information) is smaller
than 62\% (this large discrepancy is obtained for halo 10 when observed at $45^\circ$
with respect to the major axis). The next biggest error was obtained for halo 5 (37\%
when observed at $45^\circ$ with respect to the major axis).

In the case of the two remaining parameters, concentration $c$ and anisotropy $\beta$,
all haloes show common trend in the discrepancies: $\beta$ is underestimated while $c$
overestimated, especially for the fitting including all dark matter particles. The
discrepancies can be traced to the specific behaviour of the $\sigma_{\rm los}$ and
$\kappa_{\rm los}$ profiles. For example, in the case of halo 1 the three kurtosis
profiles (see Figures~\ref{mom0}-\ref{mom90}) give similar $\beta$ values, however the
$\sigma_{\rm los}$ is much shallower for the observation angle of $45^\circ$ than in the
two remaining cases. As already mentioned, $\kappa_{\rm los}$ is mainly sensitive to the
velocity anisotropy. Since the line-of-sight velocity dispersion  profile can be made
steeper in the centre either with a steeper density profile or with more radial orbits,
and since the anisotropy parameter is almost the same in all three directions, we can
expect the inferred concentration to be somewhat lower for the $45^\circ$ direction,
which is indeed the case. The situation is similar for halo 4. The kurtosis forces the
anisotropy to be very tangential, so the very steep $\sigma_{\rm los}$ profile at
$0^\circ$ (the leftmost middle panel of Figure~\ref{momrest2}) requires a very large
concentration of the density profile. 

We have calculated the means and standard deviations of the differences between the
fitted parameters ($\log M_{100}$, $\log c$, and $\beta$ or $\log
\sigma_r/\sigma_\theta$) and their values  measured from the full 6D particle phase
space.  As expected from Figure~\ref{fits}, we have found no significant dependence on
the viewing angle.  We show these quantities in Table~\ref{stats} for the 30 cases
studied (3 for each of the 10 haloes).  We can see from this Table that, with 40
particles per bin, the virial mass is a little more underestimated, while the
concentration parameter is less overestimated and the anisotropy less underestimated.
This behaviour of $c$ and $\beta$ can be understood by recalling that in the case of
using only 40 particles per bin the sampling errors of the moments are much larger. For
very  negative kurtosis small errors may enforce low $\beta$ estimates, which have to
made  up by high $c$ values (to reproduce velocity dispersion). Moreover, we find that
the error bars listed in Table~\ref{stats} are similar in both cases, which indicates
that the physical variations due to substructure and to the different shapes of haloes
dominate over the statistical noise or sampling errors.

\begin{table}
\tabcolsep 2pt
\caption[]{Results of the fitting procedure
\label{stats}}
\begin{center}
\begin{tabular}{cccccccccccc}
\hline
\hline
Particles & \multicolumn{2}{c}{$\Delta \log M_{100}$} &
& \multicolumn{2}{c}{$\Delta\log c$} &
& \multicolumn{2}{c}{$\Delta\beta$} &
& \multicolumn{2}{c}{$\Delta \log (\sigma_{r}/\sigma_{\theta})$} \\
\cline{2-3}
\cline{5-6}
\cline{8-9}
\cline{11-12}
per bin & mean & $\sigma$ &
& mean & $\sigma$ &
& mean & $\sigma$ &
& mean & $\sigma$ \\
\hline
All & --0.03 &  0.09 && 0.20 & 0.18 && --0.78 & 1.04 &&--0.12 & 0.12 \\
40  & --0.07 &  0.10 && 0.08 & 0.24 && --0.20 & 0.48 &&--0.04 & 0.11  \\
\hline
\end{tabular}
\end{center}
\end{table}

Table~\ref{stats2} shows the statistical significance of the biases on the parameters,
using the Student's $t$-statistic, which assumes Gaussian distributions of the
parameters ($\log M_{100}$, $\log c$, $\beta$ and $\sigma_r/\sigma_\theta$), and the
binomial statistic testing the distribution  of the signs of these parameters. The two
statistical tests lead to different conclusions about the bias of the parameters: the
Student's $t$-statistic always  leads to significant biases, while the binomial
statistic indicates no significant bias in concentration parameter and velocity
anisotropy when only 40 particles are used per radial bin. This difference is caused by
the skewed distributions of $\Delta\log c$, $\Delta \beta$ and $\Delta \log
(\sigma_r/\sigma_\theta)$. Although the Student's $t$-statistic has the advantage of
being more sensitive to the outliers in the distribution, it has the disadvantage of
only being valid for Gaussian parent distributions, which is not the case here. In
summary, in the case of 40 particles per radial bin, while the virial mass is biased
towards lower values, \ we cannot conclude that the concentration parameter and velocity
anisotropy are biased with the data we have.

\begin{table}
\caption[]{Statistical significances in the biases of the fitting procedure
\label{stats2}}
\tabcolsep 3pt
\begin{center}
\begin{tabular}{cllcllcllcll}
\hline
\hline
Particles & \multicolumn{2}{c}{$\log M_{100}$} &
& \multicolumn{2}{c}{$\log c$} &
& \multicolumn{2}{c}{$\beta$} &
& \multicolumn{2}{c}{$\log (\sigma_{r}/\sigma_{\theta})$} \\
\cline{2-3}
\cline{5-6}
\cline{8-9}
\cline{11-12}
per bin & \multicolumn{1}{c}{$P_t$} & \multicolumn{1}{c}{$P_{\rm b}$ } 
& & \multicolumn{1}{c}{$P_t$} & \multicolumn{1}{c}{$P_{\rm b}$} 
& & \multicolumn{1}{c}{$P_t$} & \multicolumn{1}{c}{$P_{\rm b}$} 
& & \multicolumn{1}{c}{$P_t$} & \multicolumn{1}{c}{$P_{\rm b}$} \\
\hline
All & 0.96 & 0.90 & & 1 & 1 & &    1 & 1 & & 1 & 1 \\
40  & 1 & 1    & & 0.96 & 0.71 & & 0.96 & 0.82 & & 0.99 & 0.82 \\
\hline
\end{tabular}
\end{center}
Notes: $P_t$ and $P_{\rm b}$ are the probabilities for bias using the
Student's $t$ and binomial statistics, respectively.
\end{table}

Given the range of $\Delta \log c$ seen in the right plots of Figure~\ref{fits} (see
also Table~\ref{stats}), the discrepancies noted by \citet{lm03} between their
concentration parameter for Coma and the lower values obtained by \citet{bg} and
\citet{bktm} in their kinematical analysis of stacked clusters are reduced. Assuming
that the departures in $\log c$ are independent of $c$ and $M_{100}$, we deduce that the
difference in $\log c$ between the measurement for the Coma cluster ($c=9.4$) by
\citeauthor{lm03} and the smaller concentration ($c=5.5$) found by \citeauthor{bktm} can
be accounted for in 27\% of our 30 projected haloes. On the other hand, the larger
difference in $\log c$ between the Coma measurement and the concentration found by
\citet{bg} can be accounted for in only 10\% of our projected haloes. Similarly, the
difference in $\log c$ between Coma and the value ($c = 6$) extrapolated from the values
found by \citet{bu} in their cosmological simulations, can be accounted for in 10 out of
our 30 projected haloes. Moreover, there is a non-negligible scatter in the relation
between halo concentration and mass \citep{jing,bu}, which reduces even more any
discrepancy with high concentration found for Coma.

\section{Discussion}

We studied the dynamical and kinematical properties of dark matter haloes obtained in
cosmological $N$-body simulations. First, using all the 3D information available, we
calculated their virial masses, radii, anisotropy parameters and estimated their density
profiles. Next, we obtained projected velocity moments, standard observables used to
estimate the dark matter content of virialized objects. We then fitted those `data' with
spherical models based on Jeans equations in order to reproduce the `observed' velocity
moments.

Our approach was similar to the one of \citet{tbw} who used lowest order Jeans equation
to model the velocity dispersion and find the masses of simulated haloes and compared
them to the real masses of those haloes. In addition to velocity dispersion, we used the
projected kurtosis profiles in a similar way to that applied recently by \citet{lm03} to
infer the properties of the Coma cluster. The use of kurtosis allows us to estimate the
anisotropy of the velocity distribution. Besides the virial mass and anisotropy
parameter, we fitted the concentration of the density profiles of the haloes.

Our results emphasize the difficulties in the use of higher velocity moments to infer
the properties of dark matter haloes. The kurtosis seems to be very sensitive to the
substructure and local matter flows. The discrepancies in the fitted properties of halo
4 can be traced to its peculiar mean radial velocity inside the virial radius which
amounts to $0.26 \,v_{\rm 100}$, while it is smaller than $0.1 \,v_{\rm 100}$ for most
of the remaining haloes (also in the tangential directions). We therefore confirm the
necessity of using only elliptical galaxies as tracers in the analysis of single
clusters in order to minimize the effects of infall. Another source of problems lies in
our very simple modelling of velocity anisotropy with constant $\beta$, while this
quantity really shows some radial dependence. Again, in the case of halo 4 this
dependence is rather unexpected, departing from isotropy in the very centre of the halo.

\begin{figure}
\centering
\resizebox{\hsize}{!}{\includegraphics{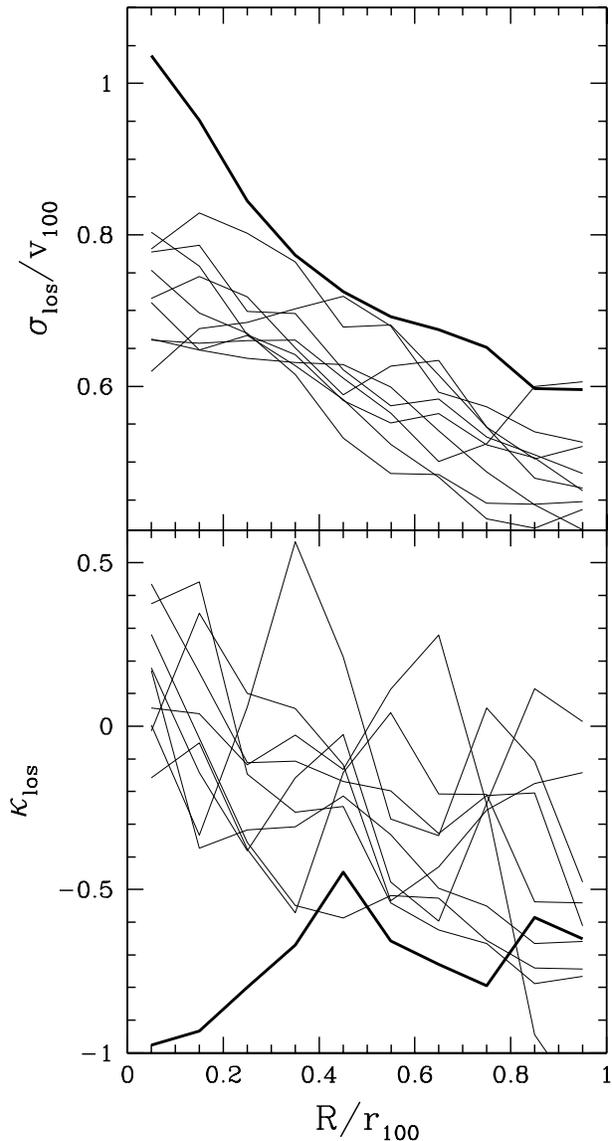}}

\caption[]{Line-of-sight velocity dispersion (\emph{top}) and kurtosis
(\emph{bottom}) vs. projected distance to the centre measured at $0^{\circ}$ to
the principal axis of the ten most massive haloes in the simulation box. The
\emph{thick line} shows the results for halo 4, while the \emph{thin lines}
show the other nine haloes.
}
\label{many}
\end{figure}

In order to check to what extent the rather discrepant results for halo~4 are an
exception or commonplace, we show, in Figure~\ref{many}, the line-of-sight velocity
dispersion and kurtosis of the ten most massive haloes in the simulation box for a
observer placed along the major axis of each halo. The thick line shows the results for
halo~4, while for the rest of the haloes thin lines are used. We infer from
Figure~\ref{many} that the velocity moments of halo 4 follow an uncharacteristic
pattern, probably caused by the unusually large negative mean radial velocity of this
halo (Table~\ref{properties}).  We have also checked the pattern of  the even velocity
moments for the other two directions of observation used in Sections~3 and~4. In these
cases, we found that the halo departing most from the general trend was halo 10, which
has the highest mean velocity inside the virial radius with respect to $v_{100}$ (with a
smaller ratio than halo 4). Therefore, local matter flows, as witnessed by non-zero mean
velocity profiles, produce perturbed line-of-sight velocity dispersion and/or kurtosis
profiles, which themselves can lead to an inaccurate estimate of the mass, concentration
and/or velocity anisotropy of a cluster of galaxies. It is important to note, however,
that concentration and anisotropy seem to be more affected than the mass estimation,
which is quite robust.

Figure~\ref{many} also shows a general trend of the behaviour of the kurtosis of the
simulated haloes. We notice that the kurtosis tends to be positive near the centre of
the halo and negative at large distances. This means that the velocity distribution is
more peaked at the centre and more flattened outside than the purely Gaussian
distribution in agreement with a recent finding of \citet{kmm03}. But in spite of these
departures from Maxwellian velocities at all radii, the Jeans analysis used here and in
\citet{lm03} produces fairly accurate measurements of the virial mass and concentration
parameter (see Fig.~\ref{fits} and Table~\ref{stats}).

Although obvious interlopers to the haloes have been removed in a similar fashion as in
\citeauthor{lm03} (\citeyear{lm03}, see also \citealp{kg82}), our results could depend
on the neighborhood of the analyzed haloes.  This would be the case if other haloes with
similar velocity were present in the direction of observation, so that their particles
would not be removed by the procedure described in Section~3. To estimate the
plausibility of such a case, we have calculated for every halo with mass up to one tenth
of the mass of the most massive halo in the simulations (that makes a total of 40
haloes) the probability that choosing an observer at a random position around it, there
would be one or more perturbing haloes with a mass greater than 25\% of the mass  of the
studied halo, at a projected distance from the halo smaller than $r_{100}$ and with a
mean line-of-sight velocity within the interval $[\overline{v}_{\rm los}-3 v_{100},
\overline{v}_{\rm los}+3 v_{100}]$, where $\overline{v}_{\rm los}$ and $v_{100}$ are the
mean line-of-sight velocity and circular velocity at $r_{100}$ of the analyzed halo.
Choosing randomly 100 observers for each halo, we found that the probability of not
finding any perturbing neighbouring halo was higher than 95\% for 93\% of the haloes, 
while for two of the analyzed haloes this probability was 88\% and only one among the 40
haloes had this probability smaller than 80\% (77\%). We can conclude from these numbers
that it is quite unlikely that the results of the Jeans analysis are affected by
neighbouring clusters. Moreover, the least isolated, i.e. the halo with the highest
probability of having a perturbing halo around,  among the top 10, halo 10, shows no
specific bias in the parameter estimates (see Fig.~\ref{fits}).

Therefore, on one hand, the observed cosmic variance of the inner structure and internal
kinematics of the massive haloes in the cosmological simulations suggests that the
typical properties of dark matter haloes are best obtained through the analysis of
stacked observations as performed by \citet*{cye97}, \citet{bg} and \citet{bktm}. But,
on the other hand, in structures (clusters of galaxies) with near zero mean velocity
profiles, this cosmic variance is much reduced. Therefore, it is well worth analyzing a
single cluster with a large number of velocities and a near zero mean velocity,  such as
was done by \citet{lm03} for the Coma cluster.

\section*{Acknowledgments}

We wish to thank the anonymous referee for his/her comments and suggestions that helped
to improve this paper and E. Salvador-Sol\'e for his comments on the manuscript. We are
also indebted to Fran\c{c}ois Bouchet, Bruno Guiderdoni and coworkers for kindly
providing us with their $N$-body simulations, especially Jeremy Blaizot for answering
our technical questions about the design and access to the simulations. TS and EL{\L}
are grateful for the hospitality of the Institut d'Astrophysique de Paris where part of
this work was done. TS and EL{\L} benefited from the Marie Curie Fellowship and the NATO
Advanced Fellowship, respectively. EL{\L} and GAM both acknowledge the France-Poland
Jumelage exchange programme. TS and GAM also acknowledge the hospitality of the
Copernicus Center in Warsaw where most of this work was performed.  Partial support was
obtained from the Polish State Committee for Scientific Research within grant No.
2P03D02319. TS acknowledges support from a fellowship of the Ministerio de Educaci\'on,
Cultura y Deporte of Spain.

\end{document}